\documentclass[a4paper,twocolumn,11pt,unpublished]{quantumarticle}\pdfoutput=1
\usepackage[utf8]{inputenc}
\usepackage[english]{babel}
\usepackage[T1]{fontenc}
\usepackage{microtype}    
\usepackage{csquotes, bm}
\usepackage{textcomp}
\usepackage{amsfonts, amsmath, amsthm, amssymb}
\usepackage[mathscr]{eucal}
\usepackage{hyperref}
\usepackage{float}
\usepackage{algorithm}
\usepackage{algpseudocode}
\usepackage{graphicx}
\usepackage{subcaption}
\usepackage[numbers,sort&compress]{natbib}
\usepackage{multirow} 
\usepackage{bbold}

\usepackage{physics}
\usepackage{nicefrac}       
\usepackage{booktabs}       
\usepackage{verbatim}       
\usepackage{xspace}
\usepackage{xr}
\usepackage{relsize}
\usepackage{enumitem}

\begin{document}
\title{Quantum Graph Attention Networks: Trainable Quantum Encoders for Inductive Graph Learning}
\author[1]{Arthur M. Faria}
\email{arthur.faria2@gmail.com}
\orcid{0000-0002-9145-7959}
\author[2]{Mehdi Djellabi}
\email{mehdi.djellabi@pasqal.com}
\orcid{0000-0003-3537-7855}
\author[1,2]{Igor O. Sokolov}
\email{i.sokolov@pasqal.com}
\orcid{0000-0002-0022-5686}
%
%
\author[1]{Savvas Varsamopoulos}
\email{svarsamo@gmail.com}
\orcid{0000-0002-5277-8768}

\affiliation[1]{Pasqal NL, Fred. Roeskestraat 100, 1076 ED Amsterdam, Netherlands}
\affiliation[2]{Pasqal SAS, 24 Av. Emile Baudot, 91120 Palaiseau, France}

\maketitle
\begin{abstract}
  We introduce Quantum Graph Attention Networks (QGATs) as trainable quantum encoders for inductive learning on graphs, extending the Quantum Graph Neural Networks (QGNN) framework introduced in~\cite{faria2025}. 
  QGATs leverage parameterized quantum circuits to encode node features and neighborhood structures, with quantum attention mechanisms modulating the contribution of each neighbor via dynamically learned unitaries. 
  This allows for expressive, locality-aware quantum representations that can generalize across unseen graph instances. 
  We evaluate our approach on the QM9 dataset, targeting the prediction of various chemical properties. 
  Our experiments compare classical and quantum graph neural networks—with and without attention layers—demonstrating that attention consistently improves performance in both paradigms. 
  Notably, we observe that quantum attention yields increasing benefits as graph size grows, with QGATs significantly outperforming their non-attentive quantum counterparts on larger molecular graphs. 
  Furthermore, for smaller graphs, QGATs achieve predictive accuracy comparable to classical GAT models, highlighting their viability as expressive quantum encoders. 
  These results show the potential of quantum attention mechanisms to enhance the inductive capacity of QGNN in chemistry and beyond.
\end{abstract}

\section{Introduction}\label{sec:intro}
Graphs are a fundamental data structure for modeling relational systems, where entities (nodes) are connected by pairwise interactions (edges). This representation naturally arises in a wide range of domains, including chemistry (molecular structures)~\cite{gircha2023}, social networks~\cite{singh2025}, Transportation \& Logistics~\cite{niu2025}, Electrical Grids \& Circuits~\cite{ullah2022}, Communication Networks, Finance and Economics~\cite{herman2022}, and many more. Traditional machine learning models often struggle to process such non-Euclidean data due to their irregular topology. To address this, Graph Neural Networks (GNNs) have emerged as a powerful class of models that learn over graph-structured inputs by iteratively aggregating and transforming information from a node’s neighborhood. By capturing both local structure and node features, GNNs enable tasks such as node classification, link prediction, and graph-level regression. Among their many variants, models like Graph Convolutional Networks (GCNs)~\cite{welling2017}, Graph Attention Networks (GATs)~\cite{veličković2018}, and GraphSAGE~\cite{leskovec2018} have demonstrated strong performance in both transductive and inductive settings, making GNNs a key building block in modern geometric deep learning.

While classical GNNs have achieved remarkable success, their scalability and expressiveness can be limited by the classical nature of their computation~\cite{abadal2021}, especially when modeling systems with inherent quantum structure, such as molecules or quantum materials. 
Quantum Graph Neural Networks (QGNNs) aim to address this by leveraging the computational power of parameterized quantum circuits to encode and process graph data in a quantum-enhanced latent space. 
By embedding node and edge features into quantum states and performing learnable unitary transformations, QGNNs can capture complex, high-dimensional correlations that may be intractable for classical models~\cite{abbas2021, havlíček2019}.

At our previous work~\cite{faria2025}, we adapted the core idea of the classical GraphSAGE framework to the quantum setting. In this architecture, each node's feature vector is encoded into a quantum state, and message passing is performed via parameterized quantum circuits that aggregate information from neighboring nodes. The quantum circuit realizes a Quantum Graph Convolutional Network (QGCN), inspired by prior quantum convolutional architectures~\cite{lukin2019}. The aggregation follows a fixed scheme, such as mean or sum, mirroring classical GraphSAGE, but the transformation and embedding of the aggregated message is executed within a quantum processor using trainable unitary operations. This quantum 
aggregation procedure is repeated across multiple layers, enabling the model to capture multi-hop dependencies in the graph while remaining inductive in nature. Although this architecture does not incorporate any dynamic weighting of neighbors, it establishes a strong and well-defined foundation for evaluating the effect of quantum-enhanced representation learning.

Attention mechanisms are fundamental in classical~\cite{vaswani2017, bahdanau2014, niu2021, veličković2018, kim2017} and quantum~\cite{li2022, kamata2025, zhao2023, widdows2024, evans2024, guo2024, gao2024, cherrat2022} machine learning, demonstrating broad effectiveness. 
Building on the QGNN framework~\cite{faria2025}, we extend our model by incorporating attention mechanisms, inspired by the effectiveness of Graph Attention Networks (GATs) in classical deep learning~\cite{veličković2018}. GATs improve graph neural networks by computing learnable, context-dependent attention coefficients that weight the influence of neighboring nodes during message aggregation. This allows the model to differentiate between structurally similar but semantically distinct neighborhoods, addressing a key limitation of standard GNNs that treat all neighbors uniformly or heuristically. In our approach however, we adapt this concept to the quantum setting by introducing Quantum Graph Attention Networks (QGATs)—an architecture that integrates self-attention mechanisms with parameterized quantum circuits. During training, QGATs dynamically learn to assign attention weights to neighboring nodes based on their features and their relationship to the target node. These attention weights are then used to modulate the behavior of trainable quantum circuits responsible for aggregating and updating node representations. By doing so, QGATs enable the quantum model to selectively emphasize informative substructures within the graph, enhancing its capacity to model complex dependencies and localized interactions. This combination of quantum processing and attention-driven adaptability makes QGATs especially effective for learning over molecular graphs, where both node-specific properties and relational context are critical for accurate prediction.

The main contributions of this work are: 
\begin{enumerate}
    \item We propose QGAT, a novel framework that augments QGNNs with attention mechanisms and implement it in \textsc{PyTorch}~\cite{pytorch} using \textsc{Qadence}~\cite{qadence2025}. In this approach, graph-structured data is embedded through tunable feature maps in quantum circuits, parameterized by attention weights.
    \item We evaluate QGAT on the QM9 molecular property prediction benchmark, reporting training results for classical GNNs and GATs alongside QGNNs and our proposed framework. 
    We consider two aggregation strategies: (i) a single shared model across all steps and (ii) step-specific models. 
    Our results demonstrate that QGNNs and QGATs can achieve predictive accuracy on par with their classical counterparts.
    \item In our numerical experiments, we show that QGATs, through the incorporation of attention mechanisms, consistently achieve higher accuracy than QGNNs on the QM9 dataset.
\end{enumerate}

The remainder of this paper is structured as follows. Section~\ref{sec:data} introduces the problem setting and describes the dataset used in our experiments. Section~\ref{sec:gnn} provides an overview of GATs. Section~\ref{sec:qgats} details the architecture of the proposed quantum counterpart, QGATs. Section~\ref{sec:res} presents the experimental results and performance analysis. Finally, Section~\ref{sec:conc} concludes the paper with a summary of our findings and discusses potential directions for future work.    

\section{Molecular Dataset: QM9}\label{sec:data}
The QM9 dataset~\cite{ramakrishnan2014quantum} is a widely used benchmark in computational chemistry for evaluating machine learning models. It was constructed from the GDB-17 database~\cite{ruddigkeit2012enumeration} and contains small organic molecules with up to nine heavy atoms such as carbon, oxygen, nitrogen, and fluorine. The dataset provides a range of properties computed using Density Functional Theory (DFT) at the B3LYP/6-31G(2df,p)~\cite{becke1993density, ditchfield1971self} level of theory, including molecular geometries, total energies, electronic properties, and thermodynamic quantities such as atomization energies, dipole moments, and HOMO–LUMO gaps. Molecular properties such as LUMO (Lowest Unoccupied Molecular Orbital) and HOMO (Highest Occupied Molecular Orbital) energies are indicators of a molecule’s chemical reactivity and stability. As these are global molecular-level properties, we frame LUMO energy prediction as a node regression task using atomic features. 

In this investigation, each molecule is represented as a graph, where nodes encode atomic properties, and edges indicate connectivity between atoms.  For each molecule in the QM9 dataset, we extract atom features and the adjacency matrix. However, only the atom features are used as classical input to the model, while the adjacency matrix specifies the set of atom features to be used.
In fact, our model uses seven atomic features: atomic number, chirality, degree, formal charge, radical electrons, hybridization state, and scaled mass (indices 0-5, 10). We omit four properties (indices 6-9: aromaticity, hydrogen count, ring membership, and valence) as their values are negligible (effectively zero) in our samples. This framework successfully connects local atomic characteristics with emergent molecular behavior.

\section{Classical Graph Neural Networks }\label{sec:gnn}
GNNs are deep learning models that operate on graph-structured data, leveraging node features and topology to learn representations. Unlike traditional neural networks, GNNs handle non-Euclidean structures. Many follow the Message Passing Neural Network (MPNN) framework, where each node aggregates information from neighbors through message and update functions over several steps. A readout function may then compute graph-level outputs. Notable models like GCNs~\cite{welling2017}, GraphSAGE~\cite{leskovec2018}, GATs~\cite{veličković2018}, and GINs~\cite{xu2018how} fall under this framework, with applications in chemistry~\cite{pmlr-v70-gilmer17a}, social networks, and recommendation systems~\cite{9693280}. 

Inspired by convolutional neural networks for image processing~\cite{Lecun_CNN}, GNNs generalize the convolution operation to graph-structured data. The layer-wise propagation rule for Graph Convolutional Networks (GCNs), introduced in~\cite{welling2017}. However, GCNs scale poorly to large graphs due to the cost of multiplying the normalized adjacency matrix with node features at each layer. They operate in a \textit{transductive} setting, requiring the full graph in memory, thus preventing mini-batch training and limiting generalization. These constraints make GCNs impractical for large-scale applications, prompting the development of more scalable models.

\begin{figure}[ht!]
    \centering
    \begin{subfigure}{0.27\textwidth}
        \centering
        \includegraphics[width=\textwidth]{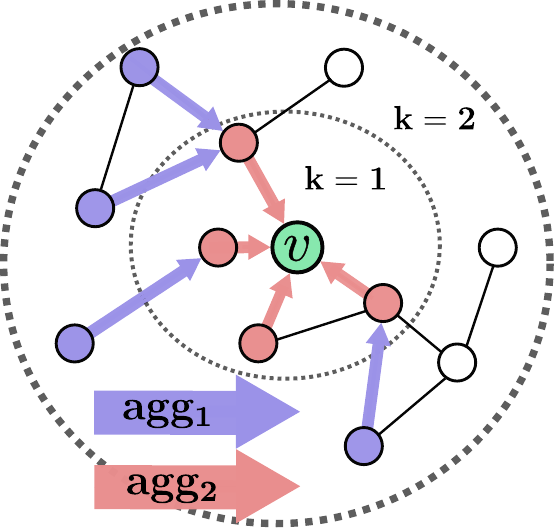}
        \caption{Aggregation steps}
        \label{fig:aggr_comp}
    \end{subfigure}
    
    \vspace{5mm}
    \begin{subfigure}{0.49\textwidth}
        \centering
        \includegraphics[width=\textwidth]{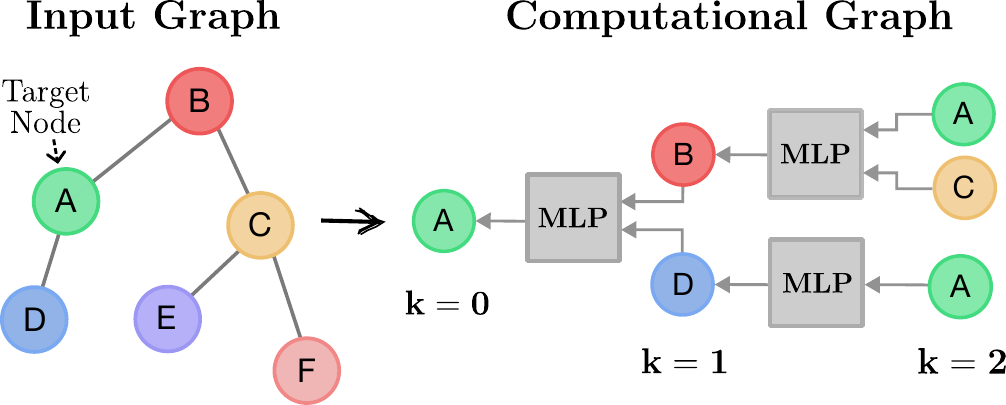}
        \caption{Computational graph scheme}
        \label{fig:comp_graph}
    \end{subfigure}
    \caption{
    Illustration of the GNN framework. (a) Message passing with aggregations (\textit{agg1}, \textit{agg2}) to the target node~$v$ over $k=2$ hops. (b) Transformation of an input graph (left) into a computation graph (right), where nodes represent feature embeddings and edges denote information flow. Gray boxes indicate Multi-Layer Perceptrons (MLPs) that aggregate neighborhood features through modular transformations.}
    \label{fig:graphsage}
\end{figure}

Among these alternatives, Ref.~\cite{leskovec2018} proposes GraphSAGE, a general inductive framework that addresses GCN limitations by enabling inference on unseen graph data. Essentially, it samples a fixed number of neighbors at each hop and applies an aggregation function to generate local embeddings, which can be concatenated with node features for richer representations. By stacking hops, the method captures multi-hop structural information while supporting mini-batching and GPU training. This sampling–aggregation scheme preserves local structure and scales to large graphs as seen in Fig.~\ref{fig:graphsage}. 
Hence, we adopt GraphSAGE as the classical GNN baseline against which to compare our quantum-enhanced models.
We next present the mathematical formulation of GraphSAGE, focusing on embedding generation and propagation.
%

\subsection{The GraphSAGE Model}
In GraphSAGE, embedding generation involves sampling and aggregation. At layer $l = 0$, the message for node $v$ is $\mathbf{h}_v^{(0)} = \mathbf{x}_v$, where $\mathbf{x}_v$ are its input features. The sampling step gathers features from neighbors $u$ of $v$ using the computational graph. For inductive learning, the message at layer $l$ for neighbor $u$ of $v$ is given by
\begin{align}
    \mathbf{h}_{\mathcal{N}(v)}^{(l)} = \textsc{Agg}^{(l)} \left( \{\mathbf{h}_u^{(l-1)}, \forall u \in \mathcal{N}(v)\} \right).
\end{align}
Here, $\textsc{Agg}$ denotes a general vector function~\cite{hochreiter1997, qi2016}. We adopt the $\textsc{Mean}$ aggregator for its simplicity and permutation invariance. It computes the element-wise mean of each neighbor node embedding $\mathbf{h}_u^{(l-1)}$, yielding $\mathbf{h}_{\mathcal{N}(v)}^{(l)} = \textsc{mean}(\{\mathbf{h}_u^{(l-1)} \mid \forall u \in \mathcal{N}(v)\})$, similar to the transductive setting~\cite{welling2017}. The target node embedding is then formed by concatenating $\mathbf{h}^{(l-1)}_v$ with $\mathbf{h}^{(l)}_{\mathcal{N}(v)}$, unlike standard GCNs. The final embedding of $v$ is given by
\begin{align}
    \mathbf{h}_v^{(l)} = \sigma \left( \mathbf{W}^{(l)} \cdot \left[ 
    \mathbf{h}_{\mathcal{N}(v)}^{(l)} \, \| \, \mathbf{h}_v^{(l-1)} 
    \right] \right)
    \label{eq:concat}
\end{align}
\noindent where $[\cdot \| \cdot]$ denotes concatenation. At the final layer $L$, the embedding $\mathbf{h}_v^{(L)}$ captures both the node’s features and its structural context. By leveraging feature information, GraphSAGE learns the node's role in the graph. This process can be enhanced with attention mechanisms~\cite{vaswani2017}, which weigh neighbors by relevance, as in ref.~\cite{veličković2018}, discussed in the next section.

\subsection{Graph Attention Networks (GATs)}
\begin{figure}[H]
    \centering
    \includegraphics[width=0.21\textwidth]{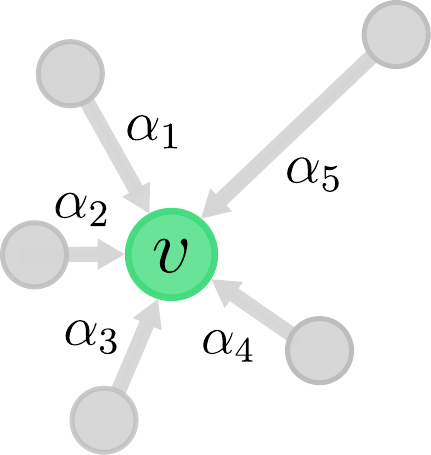}
    \caption{Schematic representation of attention weight assignment to the neighbors of a target node. Each neighboring node is assigned a weight $\alpha_1, \ldots, \alpha_7$ indicating its relative importance during message aggregation.}
    \label{fig:attent}
\end{figure}

GATs address the uniform weighting in early GNNs by introducing masked self-attention~\cite{vaswani2017}, allowing nodes to assign data-dependent importance to neighbors. This emphasizes relevant connections while down-weighting noisy ones. Attention is computed only over existing edges, retaining the parallelism and receptive-field control of Transformers, while avoiding expensive eigendecompositions. The approach scales linearly with the number of edges and supports inductive learning. Formally, for each neighbor $u \in \mathcal{N}(v)$, attention coefficients are computed as~\cite{veličković2018}
\begin{align}
    e_{vu}^{(l)} = a(\textbf{h}^{(l)}_v, \textbf{h}^{(l)}_u),
\end{align}
where the constant function $a(\cdot, \cdot)$ defines attentional mechanism between embeddings $\textbf{h}^{(l)}_v$ and $\textbf{h}^{(l)}_u$. The attention coefficients are then normalized across all neighbors of $v$ using the softmax function as
\begin{align}
    \alpha_{vu}^{(l)} = \frac{\exp\bigl(e_{vu}\bigr)}
         {\displaystyle\sum_{w \in \mathcal{N}(v)}\exp\bigl(e_{vw}\bigr)},
\end{align}
Fig.~\ref{fig:attent} schematically illustrates how attention weights are assigned to the neighboring nodes of a target node. The updated embedding of the weighted aggregation of neighbors of $v$ is hence obtained as
\begin{align}
    \mathbf{h}^{(l)}_{\mathcal{N}(v)} = \sigma\left(\sum_{u \in \mathcal{N}(u)} \alpha_{vu}^{(l)}\,\mathbf{h}^{(l-1)}_u\right)
    \label{eq:atten_weight}
\end{align}
Here again, $\sigma(\cdot)$ denotes a nonlinearity such as ReLU. The updated representation $\mathbf{h}_v^{(l)}$ is obtained by concatenating the outputs from all attention heads $\mathbf{h}^{(l)}_{\mathcal{N}(v)}$ with the self-embedding $\mathbf{h}_v^{(l-1)}$, as shown in Eq.~\eqref{eq:concat}. This extends naturally to multi-head attention, where each head uses its own projection matrix $\mathbf{W}^{(l)}$.

GATs leverage masked self-attention to down-weight noisy or high-degree nodes and provide interpretable importance scores. Since updates rely only on local neighborhoods, GATs support inductive learning across different graphs. To scale further, GraphSAGE style neighborhood sampling can be combined with GATs, forming an implicit ensemble that stabilizes training. This enables efficient mini-batching and reduces memory usage, maximizing GPU efficiency.
Next, we discuss our quantum version of the GAT model.

\section{Quantum GATs}\label{sec:qgats}
QGATs leverages quantum GraphSAGE, introduced in Ref.~\cite{faria2025}, by combining a trainable quantum feature map, serving as an attention mechanism, with a QGCN-inspired Ansatz. The feature map uses input-dependent unitaries to encode classical data into quantum states with learnable neighbor weights, which are then processed by quantum convolution and pooling layers~\cite{faria2025}.

\subsection{Trainable quantum feature map as attention mechanism}
A quantum model is trained on the basis of the classical dataset that is provided. 
The quantum model consists of a variational quantum circuit defined on a qubit register. 
The qubit register is initialized in the zero state $|0\rangle^{\otimes N}$ and a quantum feature map (FM) $\mathcal{F}$ is used to map coordinates from classical input values $\bm{x}$ to a distinct place in the Hilbert space of the qubit register. 

By assigning attention coefficients $\alpha_{vu}$ to each neighbor $u$ of a target node $v$, the classical node feature $\bm{x}_u$ is weighted by $\alpha_{vu}$. The feature map $\mathcal{F}$ is then realized by a parametrized unitary $U_{x^{\alpha_{vu}}}$ acting on the initial state $|0\rangle^{\otimes N}$. Concretely,
\begin{align}
    \bm{x}_u^{\alpha_{vu}} &\mapsto |\mathcal{F}(\bm{x}_u^{\alpha_{vu}})\rangle = U_{x^{\alpha_{vu}}}|0\rangle^{\otimes N},
\end{align}
with $\bm{x}_u^{\alpha_{vu}} \equiv \alpha_{vu}\bm{x}_u$. We refer to the state $|\mathcal{F}(\bm{x}_u^{\alpha_{vu}})\rangle$, generated by the feature map $\mathcal{F}$, as the quantum attention layer, with trainable coefficient $\alpha_{vu}$. Note that, the layer index $(l)$ is omitted here, since the weights only parametrize the FM in Eq.~\eqref{eq:atten_weight}, which has depth $1$. Hence, the initial state becomes
\begin{align}
    \rho^{\alpha_{vu}}_{u} =  |\mathcal{F}(\bm{x}_u^{\alpha_{vu}})\rangle  \langle\mathcal{F}(\bm{x}_u^{\alpha_{vu}})|
\end{align}
Trainable feature maps have previously been explored in~\cite{lloyd2020, jaderberg2024}. Unlike prior QGNNs~\cite{faria2025}, the input state $\rho^{\bm{\alpha}}$ is parametrized by trainbale attention weights $\alpha$. This formulation allows each node to adaptively attend to its neighbors by assigning distinct probabilities to their edges. This enables the Ansatz to perform flexible aggregation that naturally adapts to varying node degrees without assuming fixed or uniform connectivity. 

For a set of $N$ neighbors of node $v$, the overall input state becomes
\begin{align}
    \rho_{\mathrm{in}}^{\bm{\alpha}_v}= \bigotimes_{w=1}^N \, \rho^{\alpha_{vw}}_{w},
\end{align}
where $\bm{\alpha}_v := \{\alpha_{vw}\}_{w=1}^N$ and $\rho^{\alpha_{vw}}_{w}$ represents the state of the neighboring node $w$ of $v$ parametrized by the corresponding attention weight $\alpha_{vw}$. The feature map $\mathcal{F}$ is applied uniformly across all neighbors.

\subsection{Ansatz-QGCN circuit}
\begin{figure}[ht!]
    \centering
    \begin{subfigure}{0.34\textwidth}
        \centering
        \hspace{-10mm}\includegraphics[width=\textwidth]{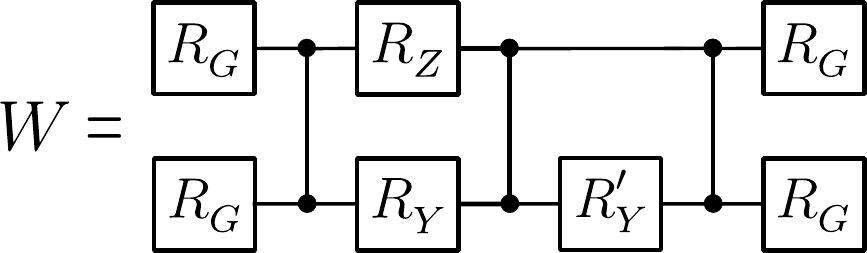}
        \vspace{2mm}
        \caption{$W$ definition.}
    \label{fig:ansatz}
    \end{subfigure}
    
    \vspace{3.mm}
    \begin{subfigure}{0.48\textwidth}
        \centering
        \includegraphics[width=\textwidth]{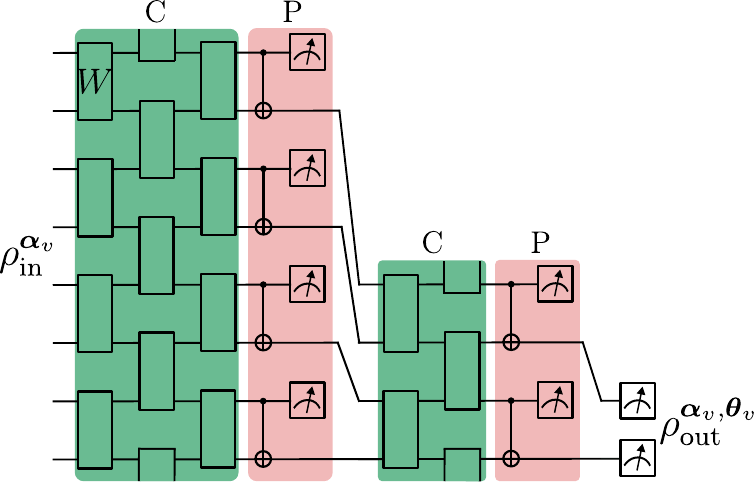}
        \caption{QGCN circuit.}
        \label{fig:qgnn_circuit}
    \end{subfigure} 
    \caption{
    Schematic representation of the QGCN circuit and convolutional unitary $W$ acting on neighboring qubits. (a) Definition of the convolutional cell $W$, which consists of CZ gates applied to alternating qubit pairs, enclosed by parameterized single-qubit rotations $R_G$. The single qubit gates: $R_Z, R_Y$, and, $R'_Y$ also are parametrized with different tunable angles. (b) The QGCN processes an $n$-qubit input state $\rho_{\mathrm{in}}^{\bm{\alpha}_v}$ through $L$ layers, each consisting of alternating convolution $(\mathrm{C})$ and pooling $(\mathrm{P})$ operations ($L = 2$ in this case). The $\mathrm{C}$ layers apply unitary transformations $W$ to qubit pairs and $\mathrm{P}$ layers reduce dimensionality by tracing out half the qubits, ultimately yielding the output state $\rho_{\mathrm{out}}^{\bm{\alpha}_v, \bm{\theta}_v}$.}
    \label{fig:qgcn}
\end{figure}

Inspired by Ref.~\cite{lukin2019, faria2025}, the QGCN circuit is described as a composition of quantum convolutional and pooling channels as
\begin{align}
     \Phi_{\bm{\theta}_v} =
     \bigcirc_{l=1}^{L} \left(\mathrm{P}_{l} \circ \mathrm{C}_{l}^{\bm{\theta}^{l}_v}\right),
     \label{eq:qgcn-map}
\end{align}
\noindent where $\bm{\theta}_v := \{\bm{\theta}^{l}_v\}_{l=1}^L$ and $\circ$ denotes a single function composition, and $\bigcirc_{l=1}^L$ represents the composition of $L$ functions applied sequentially. Each QGCN layer $l$ comprises a quantum convolutional layer $\mathrm{C}_l^{\bm{\theta}^{l}}$, with $\bm{\theta}^{l}$ being the convolution parameters, followed by a quantum pooling layer $\mathrm{P}_{l}$. The alternating structure of the QGCN circuit processes and simplifies the input quantum state, starting with $\mathrm{C}_1$ and ending with $\mathrm{P}_{L}$. 
The convolutional layer acts on an even number of qubits and preserves the size of the quantum register defined as
\begin{align}
    \mathrm{C}_l^{\bm{\theta}^l}(\cdot) \hspace{-0.5mm}=\hspace{-0.5mm} 
    \scalebox{1}{$\bigcirc$}_{j=1}^{r} \hspace{-1.5mm} \left(\hspace{-0.5mm}\bigotimes_{(i,i+1)  \in \mathrm{S}(j)} \hspace{-0.5mm}W_l^{(i,i+1)}\hspace{-1mm}\left(\bm{\theta}^l\right)\right)(\cdot),
    \label{eq:conv_layers}
\end{align}
\noindent with $W_l$  processing adjacent qubit pairs $(i, i+1)$, as displayed in Fig.~\ref{fig:ansatz}. For $n_l$ qubits in the layer $l$, the operator $W_l$ acts on nearest-neighbor pairs in an alternating fashion: at even steps $j$, it applies to pairs $(0,1), (2,3), \dots$; at odd steps $j$, it applies to $(1,2), (3,4), \dots$. Formally, the pairs are $S(j) = \{(i, i+1) \,\mid\, i \equiv j \pmod{2},\, 0 \leq i \leq n_l-2 \}$. Using the decomposition proposed in Ref.~\cite{vatan2004}, $W_l$ is composed of 3 CZ gates, 3 single-qubit rotations, and 4 general $R^G$ gates, where
\begin{align}
    R^G(\bm{\theta}^l) = e^{-iX\theta_1^{l}/2}e^{-iZ\theta_2^{l}/2}e^{-iX\theta_3^{l}/2}.
\end{align}
Each $\mathrm{C}_l$ layer has depth $r$ (e.g., $r_1$, $r_2$ in a two-layer QGCNs). Conversely, the pooling layer $\mathrm{P}_{l}$ reduces the size of the quantum register by tracing out selected qubits as
\begin{align}\label{eq:pooling-layer}
    \mathrm{P}_{l} = \Tr_{i}[(\cdot)],
\end{align}
where $i$ denotes the qubits traced out at layer $l$. In practice, this discards half of the qubits at each pooling step, i.e., $i = \{ j \in \{0,1,\dots,n_l-1\} \,|\, j \equiv 0 \pmod{2} \}$. Generally, the reduction is implemented by measuring a subset of qubits and applying classically controlled Pauli gates on their neighbors. As a design choice, the principle of deferred measurement~\cite{nielsen} is applied. Thus, the pooling layers first employ controlled operations, while (partial) measurements are postponed until the end of the computation. At the final QGCN layer $L$, the output state produced by the quantum aggregation, $\rho_{\mathrm{out}}$, is defined as
\begin{align}
    \rho_{\mathrm{out}}^{\bm{\alpha}_v, \bm{\theta}_v} &= \Phi_{\bm{\theta}_v}(\rho^{\bm{\alpha}_v}_{\mathrm{in}})
    \\
    &=\bigotimes_{w=1}^N \, \Phi_{\bm{\theta}_v} (\rho^{\alpha_{vw}}_{\mathrm{in}})
\end{align}
For the full circuit architecture, see Fig.~\ref{fig:qgnn_circuit}. 

\subsection{Embedding generation and model training}
\begin{figure}[ht!]
    \centering
    \includegraphics[width=0.42\textwidth]{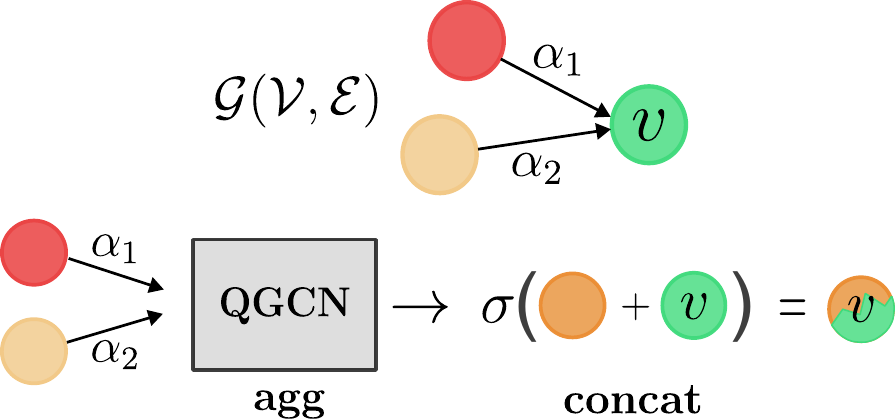}
    \caption{Schematic of the proposed QGAT framework using a QGCN circuit as quantum aggregator, which contains $L$ sets of $\mathrm{C}$ and $\mathrm{P}$ layers. Given a graph $\mathcal{G}(\mathcal{V},\mathcal{E})$, node features from connected nodes (red, yellow) are sampled and processed by the QGCN to generate the neighbor embeddings (orange) $\mathbf{h}_{\mathcal{N}(v)}^{(L)}(\alpha_1, \alpha_2, \bm{\theta}_v)$. These are concatenated classically with the target node’s self-embedding (green) through a simple vector operation and transformed by a non-linear function $\sigma$ to yield the updated embedding $\mathbf{h}_v^{(L)}(\alpha_1, \alpha_2, \bm{\theta}_v)$.}
    \label{fig:graph_fram}
\end{figure}

By employing the trainable FM to initialize the weighted qubit states and the QGCN as the aggregator, the classical embedding for the neighbors $u$ of $v$ is obtained through measurements of the observable ${\mathcal{O}}$, expressed as
\begin{align}
    \mathbf{h}_{\mathcal{N}(v)}^{(L)}(\bm{\alpha}_v, \bm{\theta}_v) &= \Tr[\rho_{\mathrm{out}}^{\bm{\alpha}_v, \bm{\theta}_v} {\mathcal{O}}]
    \\
    &= \langle {\mathcal{O}}\rangle_{\rho_{\mathrm{out}}^{\bm{\alpha}_v, \bm{\theta}_v}} .
\end{align}
Note that the embedding is generated by the same QGCN regardless of the number of neighboring nodes $u$. It can be further concatenated with the self-embedding of node $v$ to yield its updated representation as
\begin{align}
    \mathbf{h}_v^{(L)}(\bm{\alpha}_v, \bm{\theta}_v) = \sigma \left(\left[ 
    \mathbf{h}_{\mathcal{N}(v)}^{(L)} \, \| \, \mathbf{h}_v^{(1)} \right] \right).
    \label{eq:concat2}
\end{align}
As the QGCN circuit has $L$ layers, the initial embedding $\mathbf{h}_v^{(1)}$ can be only concatenated after the quantum measurement. In addition, the concatenation operation used here is a simple vector operation that does not involve any learnable parameters.  Fig.~\ref{fig:graph_fram} summarizes the proposed QGAT framework.

For completeness, in a multi-head setting, where multiple target nodes $\bm{\nu} := \{v_k\}_{k=1}^K$ are considered, the output embedding reads
\begin{align}
    \mathbf{h}_{\bm{\nu}}(\bm{\alpha}, \bm{\theta}) = \sigma\left(\left[\bigg\|_{k=1}^{K} \mathbf{h}_{v_k}^{(L)}\right]\right),
\end{align}
where $\bm{\theta} := \{ \bm{\theta}_{v_k} \}_{k=1}^{K}$ and $\bm{\alpha} := \{ \bm{\alpha}_{v_k} \}_{k=1}^K$. For simplicity, we assume target nodes have the same number of neighbors and that all QGCN Ansätze employed have a uniform depth $L$ across hops.

To this point, we have focused on the unsupervised node-level task within the graph structure. By repeatedly propagating information until reaching the designated target node, the resulting embedding captures both the intrinsic features of the node and its structural dependencies within the graph. The final embedding yields the predicted graph-level output $\bar{y}(\widetilde{\bm{\theta}}_v)$, where $\widetilde{\bm{\theta}}_v = \{\bm{\alpha}_v, \bm{\theta}_v\}$ represents the tunable parameters of the QGAT model. This allows us to run a supervised learning, by comparing the prediction to the true target value $y$. The training error is quantified by the loss function $\mathcal{L}$, and the parameters $\widetilde{\bm{\theta}}_v$ are optimized by minimizing $\mathcal{L}(\bar{y}(\widetilde{\bm{\theta}}_v), y)$ as
\begin{align}
    \widetilde{\bm{\theta}}^*_v &= \arg\min_{\widetilde{\bm{\theta}}_v} \mathcal{L}(y(\widetilde{\bm{\theta}}_v), y).
\end{align}
The parameters are efficiently minimized using standard gradient-based optimization algorithms. Successful training results in a negligible prediction error, i.e., \(|\bar{y}(\widetilde{\bm{\theta}}_v) - y| \approx 0\). We next benchmark our proposed framework on the QM9 dataset, evaluating its performance relative to its classical counterpart in molecular property prediction.

\section{Results}\label{sec:res}
We now present the key findings from our implementation. Our evaluation compares the performance of quantum-enhanced graph neural networks (QGNNs, QGATs) against their classical counterparts (GNNs, GATs) across molecules of varying sizes. 
We use GNN to refer to the GraphSAGE model.

\begin{table*}[ht!]
    \centering
    \begin{tabular}{lcccccc}
        \toprule
        Model & samples & inputs/qubits &  FM &   $r$/hidden layers & params \\
        \midrule
        Classical          & 30 & 8 &   -      & $[8]$     & 209 \\
        Quantum            & 30 & 8 & Fourier  & $[3,1,1]$ & 218 \\
        Classical (multi)  & 40 & 8 &   -      & $[2,2]$   & 1675 \\
        Quantum (multi)    & 40 & 8 & Fourier  & $[1]$     & 1700 \\
        \bottomrule
    \end{tabular}
    \caption{The table presents the hyperparameters of the classical and quantum models. For each model, we have: number of training samples, number of input features (classical) or qubits (quantum),  feature map type (quantum only), either the $r$ configuration (quantum) or hidden layer architecture (classical), and  total trainable parameters (params). The "multi" suffix denotes models combining multiple sub-models, e.g., quantum (multi) uses separate circuits for each graph hop, and similarly for classical (multi). Sample sizes were determined by model complexity: single-model architectures used $30$ samples, whereas multi-model configurations employed $40$.}
    \label{tab:comparison}
\end{table*}

\subsection{Training details}
The performance of each quantum and classical model is evaluated using Smooth L1 Loss and the $R^2$ score in separate experiments on the same small subset of molecular samples. The Smooth L1 Loss is selected for its robustness to outliers and stability, particularly advantageous for small, high-variability datasets. Training employs the Adam optimizer ($\beta_1 = \beta_2 = 0.9$) with a decaying learning rate initialized at $0.03$. 

Molecules are processed in two stages: (i) an unsupervised atom-level aggregation, where the model aggregates connected node features according to the adjacency matrix of the molecule to form a molecular representation, and (ii) a supervised molecule-level step, where this representation is used to predict the target value. Here, the loss function compares predictions with true targets, guiding the training of the model to capture both local and global structural information. Algorithm~\ref{alg:gnn_attention} presents the high-level implementation of the previous steps. In the case of the quantum models $\mathcal{M}$, the FM and Ansatz are incorporated directly into their definition

\vspace{0.5mm}
{\footnotesize
\begin{algorithm}[H]
\caption{(Q)GATs on the QM9 dataset}
\label{alg:gnn_attention}
\begin{algorithmic}[1]
\Require Batch of molecules $G$ with adjacencies $A$, atom features $F$, targets $T$, model $\mathcal{M}$, attention weights $\alpha$
\Ensure $T_\mathrm{true}$, $T_\mathrm{pred}$, $loss$
\State $loss \gets 0.0$, $T_\mathrm{pred} \gets [\,]$, $T_\mathrm{true} \gets [\,]$
\For{\textbf{each} mol $(A, F, T, \alpha)$ in $\texttt{batch}$}
    \State $prev\_out \gets \mathbf{0}$, $atom\_outs \gets [\,]$
    \For{\textbf{each} node $v$ in $G$}
        \State $a\_feat \gets F[A[v]] \times \alpha[v]$
        \State $x \gets \texttt{concat}(a\_feat, prev\_out)$
        \State $out \gets \mathcal{M}(x)$
        \State $prev\_out \gets out$, 
        \State Append $out$ to $atom\_outs$
    \EndFor
    \State $mol\_out \gets \texttt{mean}(atom\_outs)$
    \State Append $mol\_out$ to $T_\mathrm{pred}$
    \State Append $T$ to $T_\mathrm{true}$
    \State $loss \mathrel{+}= \texttt{criterion}(mol\_out, T) / |\texttt{batch}|$
\EndFor
\State \Return $T_\mathrm{true}$, $T_\mathrm{pred}$, $loss$
\end{algorithmic}
\end{algorithm}
}
\vspace{0.5mm}

In the proposed algorithm, all hops $k$ up to the number of atoms are considered; for instance, a 9-atom molecule undergoes 9 hops, with embeddings propagated from lower- to higher-index atoms, yielding a single-branch computational graph with one final head. Classical GNNs and GATs are implemented in \textsc{PyTorch}~\cite{pytorch}, while QGNNs and QGATs use \textsc{Qadence}~\cite{qadence2025}. The number of inputs/qubits are 8, matching the 7 node feature dimension plus one (accounting for the mean self-embedding of the target node)\footnote{Features such as atomic number, chirality, degree, formal charge, radical electrons, hybridization, and scaled mass are kept, while aromaticity, hydrogen count, ring membership, and valence are excluded as they are consistently zero in the selected samples.}. QM9 data are encoded via either a trainable or standard Fourier Feature Map~\cite{schuld2021, kyriienko2021}, depending on attention layers usage. Measurements assess local magnetization with tunable parameters $\bm{\omega} := \{ \omega_{i} \}_{i=1}^{N}$, and the average total magnetization is defined as an expectation value:
\begin{align}
    \langle {\mathcal{O}}\rangle = \frac{1}{N}\sum_{i=1}^N \omega_i \langle Z^{(i)}\rangle.
\end{align}

The hyperparameters of the quantum and classical models used in the results are detailed in Table~\ref{tab:comparison}. Unless otherwise specified, the models hyperparameters and evaluation metrics: Smooth L1 Loss and $R^2$ score, remain unaltered.

\subsection{Discussion}
We evaluate the performance of four graph learning frameworks: GNNs, GATs, QGNNs, and our proposed QGATs, on subsets of the QM9 dataset grouped by molecular size: molecules with up to 9, 16, 20, and 25 atoms. For each subset, we randomly sample 30/40 molecules and report the average performance in terms of training loss and $R\textsuperscript{2}$ score. Two aggregation strategies are considered: (i) using a single shared model across all aggregation steps (or hops), and (ii) employing a distinct model at each hop, thereby increasing the modularity of the architecture. Table~\ref{tab:performance} presents the complete results for all frameworks under both aggregation strategies and across the defined molecular size regimes. 

\begin{figure}[ht!]
    \centering
    \begin{subfigure}{0.49\textwidth}
        \centering
        \includegraphics[width=\textwidth]{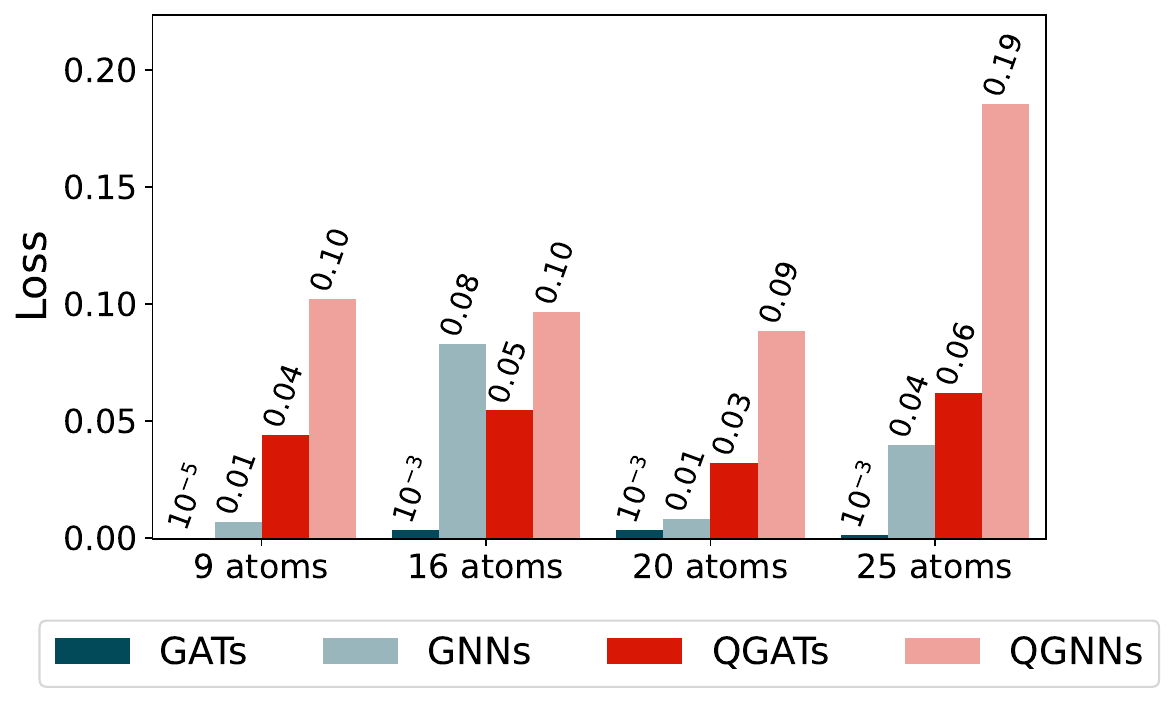}
        \caption{Loss.}
        \label{fig:single_loss}
    \end{subfigure}
    \begin{subfigure}{0.49\textwidth}
        \centering
        \includegraphics[width=\textwidth]{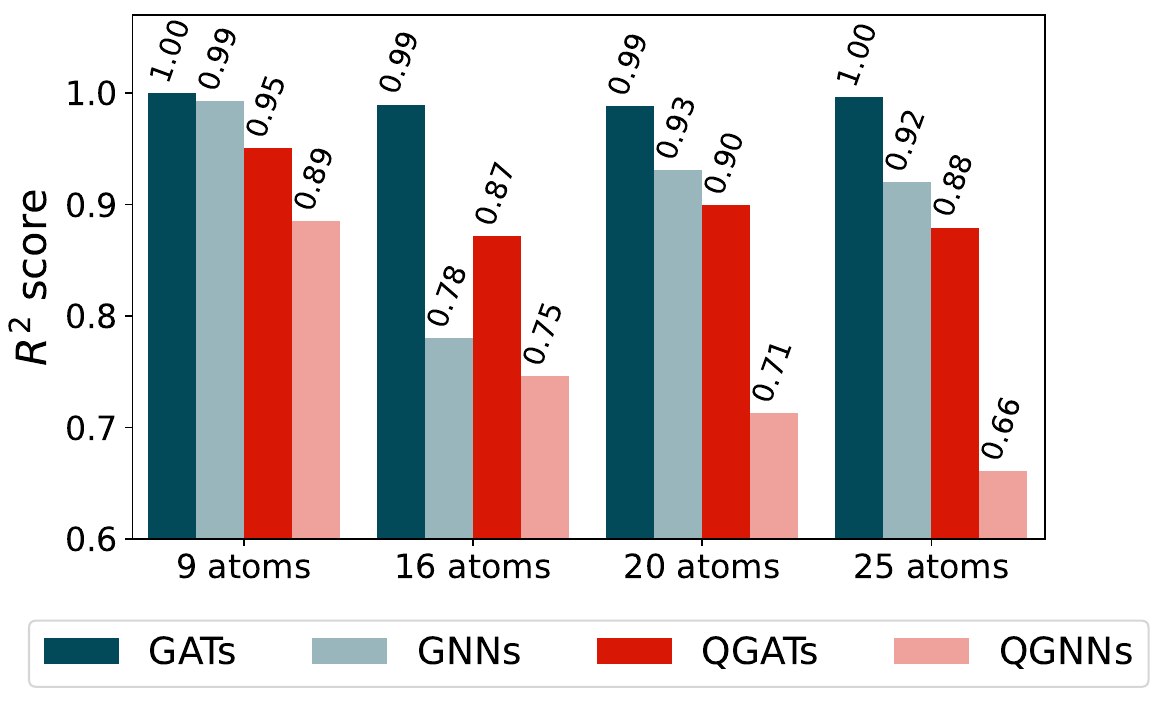}
        \caption{$R^2$ score.}
        \label{fig:single_score}
    \end{subfigure}
    \vspace{-2mm}
    \caption{Training results of single-model approach across different graph neural network architectures (GNNs, QGNNs, GATs, QGATs). The subfigure~(a) shows loss values whereas (b) R² scores for molecules of varying sizes: from $\leq 9$ to $\leq 25$ atoms.}
    \label{fig:single}
\end{figure}

The following analysis examines performance across frameworks and highlights the benefits of attention mechanisms and trainable FMs, particularly in quantum contexts. In the single-model setting, where a single GCN or QGCN is reused across all aggregation steps, classical models consistently outperform quantum ones. As shown in Figure~\ref{fig:single}, GATs achieve near-perfect scores across all molecule sizes, with GCNs also performing strongly, albeit with a slight performance drop as molecule size increases. Quantum models, by contrast, exhibit a more noticeable decline in performance as molecular complexity increases. QGCNs, in particular, struggle to generalize under this configuration. In contrast, QGATs consistently outperform QGCNs across all size regimes. This gap highlights the value of attention mechanisms in both classical and, more importantly, quantum graph learning models. The ability to adaptively weigh neighbor contributions appears to guide the quantum encoding process more effectively by emphasizing structurally or chemically relevant information.
\begin{table*}[ht!]
    \centering
    \begin{tabular}{l|ccccccccc}
        \toprule
        \multicolumn{1}{l}{} & \multicolumn{1}{l}{} & \multicolumn{2}{c}{\textbf{GNNs}} & \multicolumn{2}{c}{\textbf{QGNNs}} & \multicolumn{2}{c}{\textbf{GATs}} & \multicolumn{2}{c}{\textbf{QGATs}} 
        \\
        \cmidrule(lr){3-4} \cmidrule(lr){5-6} \cmidrule(lr){7-8} \cmidrule(lr){9-10}
        \multicolumn{1}{l}{}& \textbf{Metric} & single & multi & single & multi & single & multi & single & multi 
        \\
        \midrule
        \multirow{2}{*}{$\leq 9$} & Loss & 0.00676 & 0.00009 & 0.1021 & 0.00032 & 0.00003 & 0.00005 & 0.0441 & 0.00008
        \\
        & R² score & 0.99280 & 0.99980 & 0.8853 & 0.99840 & 1.00000 & 0.99920 & 0.9504 & 0.99990
        \\
        \addlinespace
        \multirow{2}{*}{$\leq 16$} & Loss & 0.08291 & 0.00001 & 0.0966 & 0.02753 & 0.00325 & 0.00002 & 0.0544 & 0.00398
        \\
        & R² score & 0.78000 & 1.00000 & 0.7463 & 0.93700 & 0.98960 & 0.99990 & 0.8717 & 0.98090
        \\
        \addlinespace
        \multirow{2}{*}{$\leq 20$} & Loss & 0.00810 & 0.00001 & 0.0882 & 0.03209 & 0.00337 & 0.000003 & 0.0321 & 0.00002
        \\
        & R² score & 0.93080 & 0.99970 & 0.7127 & 0.86760 & 0.98800 & 1.00000 & 0.8992 & 0.99980
        \\
        \addlinespace
        \multirow{2}{*}{$\leq 25$} & Loss & 0.03968 & 0.00003 & 0.1854 & 0.04720 & 0.00118 & 0.00001 & 0.0620 & 0.00610
        \\
        & R² score & 0.91990 & 0.99970 & 0.6610 & 0.92790 & 0.99670 & 0.99990 & 0.8793 & 0.98920
        \\
        \bottomrule
    \end{tabular}
    \caption{Comparison of single-model versus multi-model approaches across different graph neural network architectures (GNNs, QGNNs, GATs, QGATs). The table shows loss values and R² scores for molecules of varying sizes: from $\leq 9$ to $\leq 25$ atoms.}
    \label{tab:performance}
\end{table*}
\begin{figure}[ht!]
    \centering
    \begin{subfigure}{0.49\textwidth}
        \centering
        \includegraphics[width=\textwidth]{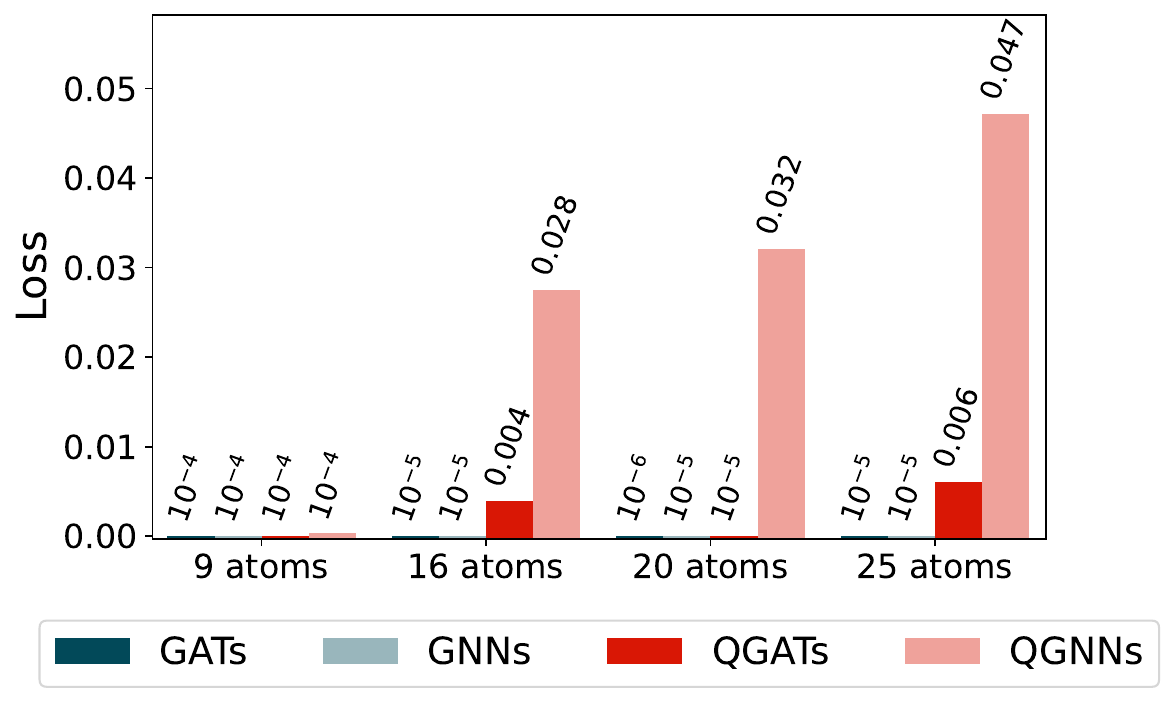}
        \caption{Loss.}
        \label{fig:multiple_loss}
    \end{subfigure}
    \begin{subfigure}{0.49\textwidth}
        \centering
        \includegraphics[width=\textwidth]{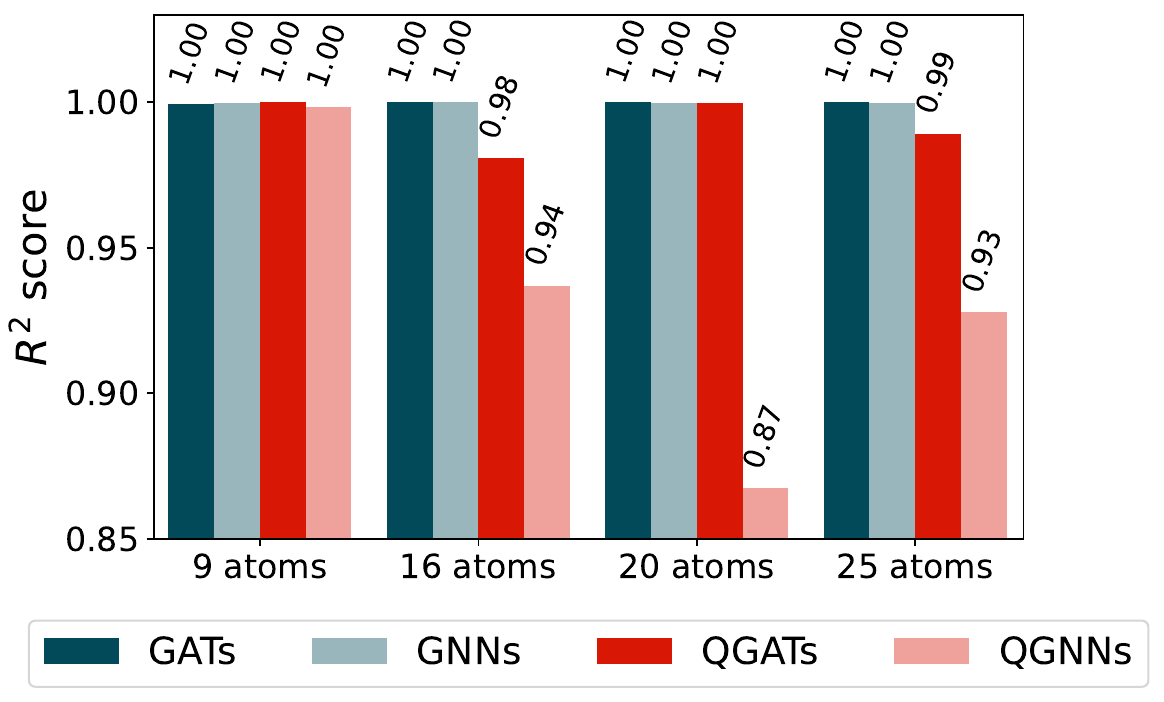}
        \caption{$R^2$ score.}
        \label{fig:multiple_score}
    \end{subfigure}
    \vspace{-2mm}
    \caption{Training results of multi-model approach across different graph neural network architectures (GNNs, QGNNs, GATs, QGATs). The subfigure~(a) shows loss values and the subfigure~(b) displays R² scores for molecules of varying sizes: from $\leq 9$ to $\leq 25$ atoms.}
    \label{fig:multiple}
\end{figure}

In the multiple-model setting, where each aggregation step is assigned a distinct, shallow classical or quantum models, we observe substantial performance improvements across all architectures. As illustrated in Figure~\ref{fig:multiple}, both classical GATs and GNNs achieve $R^2$ scores of $~1.0$, not showing a big improvement in the classical scenario. However, the quantum frameworks again experience the most significant gains from this modular approach. QGNNs, which previously underperformed, show now strong improvements across all molecule sizes. QGATs also benefit substantially, confirming the synergy between attention mechanisms and modular quantum design. Hence, by employing shallow circuits that are less prone to barren plateaus and easier to optimize, it can enhance both the stability and efficiency of the training process in quantum graph machine learning.

\section{Conclusion}  \label{sec:conc}
This work introduced QGAT, a quantum graph attention framework that integrates attention mechanisms into quantum graph neural networks for inductive learning via tunable feature maps, where the quantum circuit scales independently of the number of nodes.
Through extensive experiments on the QM9 dataset, we demonstrated that attention not only improves classical models but is particularly beneficial in the quantum setting, where it enhances expressivity and task adaptability, enabling QGATs to surpass our QGNN architectures.
Furthermore, we showed that modular architectures employing multiple shallow circuits per aggregation step significantly outperform single deep-circuit designs, offering a practical solution to the optimization challenges of quantum models. 
Together, these findings highlight a promising path toward more scalable and trainable quantum graph learning models.

\bibliographystyle{unsrt} 
\bibliography{qgats} 

\end{document}